\documentclass[aps,prc,twocolumn,showpacs,superscriptaddress,longbibliography,nofootinbib,floatfix,10pt]{revtex4-1}
\usepackage{CJK}
\usepackage{dcolumn}
\usepackage{amssymb}
\usepackage{amsmath}
\usepackage{mathrsfs}
\usepackage{graphicx}
\usepackage[english]{babel}
\usepackage{bm}
\usepackage{xcolor}
\usepackage{fix-cm}
\usepackage{mathptmx}
\usepackage[T1]{fontenc}
\usepackage[colorlinks,citecolor=blue,linkcolor=blue,anchorcolor=blue,filecolor=blue,urlcolor=blue]{hyperref}
\usepackage{multirow}

\usepackage{epstopdf}
\makeatletter
\def\NAT@def@citea{\def\@citea{\NAT@separator}}
\makeatother

\begin{document}
\begin{CJK*}{GBK}{}

\title{Microscopic studies of production cross sections in multinucleon transfer reaction $^{58}$Ni+$^{124}$Sn}

\author{Zhenji Wu}
\affiliation{School of Nuclear Science and Technology, University of Chinese Academy of Sciences, Beijing 100049, China}
\author{Lu Guo} \email{luguo@ucas.ac.cn}
\affiliation{School of Nuclear Science and Technology, University of Chinese Academy of Sciences, Beijing 100049, China}
\affiliation{Institute of Theoretical Physics, Chinese Academy of Sciences, Beijing 100190, China}

\date{\today}

\begin{abstract}
\edef\oldrightskip{\the\rightskip}
\begin{description}
        \rightskip\oldrightskip\relax
        \setlength{\parskip}{0pt}
\item[Background] Multinucleon transfer reaction at low-energy collisions is considered to be a promising method for the production of new exotic nuclei, which are difficult to be produced by other methods. The theoretical studies are required to provide reliable predictions for the experiments and promote the understanding of the microscopic mechanism in multinucleon transfer reactions.
\item[Purpose] We provide a predictive approach for production cross sections, and testify how and to what extent
the microscopic approach works well in multinucleon transfer reaction.
\item[Methods] We employ the approach TDHF+GEMINI, which combines the microscopic time-dependent Hartree-Fock (TDHF) with the state-of-art statistical model \texttt{GEMINI++}, to take into account both the multinucleon transfer dynamics and the secondary deexcitation process. The properties
    of primary products in multinucleon transfer process, such as transfer probabilities and primary
    cross sections, are
    extracted from TDHF dynamics by using the particle-number projection method. Production cross sections for secondary products are evaluated by using the statistical model \texttt{GEMINI++}.
\item[Results] We investigate the influence of colliding energies and deformation orientations of target and projectile nuclei on multinucleon transfer dynamics in the reaction $^{58}$Ni+$^{124}$Sn. More nucleons are observed to transfer in the tip collision
as compared to the side collision. The production cross sections for secondary fragments with TDHF+GEMINI calculations well reproduce the experimental measurements at energies close to the Coulomb barrier.
At sub-barrier energy, the theoretical results gradually deviate from the experimental data as the increase of the
number of transferred neutrons, implying the limitations of a single mean-field approximation in TDHF approach. Possible origins for this discrepancy are discussed. The total cross sections integrated over all the neutron pickup channels are in good agreement with the experimental data for all the energies. We compare the production cross sections of TDHF+GEMINI calculations with those from GRAZING model, and find that our approach gives a quantitatively good description as the semiclassical model, although there is no adjustable parameters for the reaction dynamics in the
microscopic TDHF method.
\item[Conclusions] The microscopic approach TDHF+GEMINI reasonably reproduces the experimental data at energies close to the Coulomb barrier and well accounts
for the multinucleon transfer mechanism.
The present studies clearly reveal the applicability of TDHF+GEMINI method in multinucleon transfer reactions, which thus deserves as a promising tool to predict the properties of new reactions.

\end{description}
\end{abstract}
\maketitle
\end{CJK*}

\section{INTRODUCTION}

The production of new exotic nuclei in both experimental and theoretical studies is of fundamental importance
to enrich our knowledge of the characteristics of the atomic nuclei, in particular for those neutron-rich isotopes involved in the
astrophysical r-process. These nuclei may show distinct properties from those seen in typical stable nuclei, which are extremely interesting for the nuclear structure and reaction mechanism investigation. In recent years
the multinucleon transfer reactions occurring in low-energy collisions of heavy ions are considered as an effective method
for the production of exotic nuclei located far from stability line.
The experiments to produce these exotic nuclei have been extensively performed~\cite{Kozulin2012_PRC86-044611,Barrett2015_PRC91-064615, Vogt2015_PRC92-024619,Watanabe2015_PRL115-172503,Devaraja2015_PLB748-199,Mijatovic2016_PRC94-064616,Welsh2017_PLB771-119,Desai2019_PRC99-044604}, providing the important information concerning
the mechanisms of multinucleon transfers in low-energy collisions. For example, the neutron-rich nuclei around $N=126$ have been produced
via the multinucleon transfer reactions~\cite{Barrett2015_PRC91-064615,Vogt2015_PRC92-024619,Watanabe2015_PRL115-172503}, which are difficult to be produced by other methods so far.
It was found that the shell effect plays an important role in the production of these neutron-rich nuclei and may significantly enhance
the yield of exotic nuclei for an appropriate projectile-target combination.
To produce the new unstable isotopes experimentally, the optimal incident energy and projectile-target combination should be
chosen to have the highest product cross section for the desired isotope. The reliable theoretical predictions are therefore
required to guide the current and future experiments at radioactive-ion beam facilities.

Various theoretical models, including both
the semiclassical and microscopic approaches, have been developed to describe the multinucleon transfer process.
The semiclassical models, such as GRAZING~\cite{Winther1994_NPA572-191}, Complex Wentzel-Kramers-Brillouin (CWKB)~\cite{Vigezzi1989_AoP192-432}, dinuclear system (DNS)~\cite{Antonenko1995_PRC51-2635,Feng2009_PRC80-057601,Adamian2010_PRC81-057602,
Wang2012_PRC85-041601,Zhu2016_PRC93-064610,Bao2016_PRC93-044615,Feng2017_PRC95-024615,Zhu2017_PLB767-437},
and the dynamical model based on Langevin-type
equations of motion~\cite{Zagrebaev2008_PRL101-122701,Zagrebaev2013_PRC87-034608,Zagrebaev2014_PRC89-054608,Karpov2017_PRC96-024618,Saiko2019_PRC99-014613} have shown remarkable successes in
reproducing the particular aspects of experimental data.
However, the uncertainty of macroscopic parameters and the lack of microscopic origins restrict their
predictive power and may obscure the underlying physical processes.
On the other hand, the microscopic approaches, e.g., improved quantum molecular dynamics (ImQMD) \cite{Wang2002_PRC65-064608,Wen2013_PRL111-012501,Wen2014_PRC90-054613,Zhao2015_PRC92-024613,Zhao2016_PRC94-024601,Li2016_PRC93-014618,
Li2018_PLB776-278} and time-dependent Hartree-Fock (TDHF) model \cite{Simenel2012_EPJA48-152,Nakatsukasa2016_RMP88-045004,Simenel2018_PPNP103-19,Stevenson2019_PPNP104-142,Sekizawa2019_FP7-20} have also been proposed to describe
the multinucleon transfer processes. These models are based on the mean-field approximation and treat the nucleon transfer and dissipation
dynamics in a self-consistent way. In ImQMD model, besides the time evolution of mean field, the stochastic two-body collision is included
in the Hamiltonian equation of motion so that the dynamical fluctuation correlation can be treated in heavy-ion collision such as the multifragmentation process. However, the omission of spin-orbit interaction in ImQMD model prevents a proper treatment of shell effects in heavy-ion collisions.
Comparing with the molecular dynamics simulations, TDHF theory has the advantage in describing better the structure effects of nuclear system such as the shell effects and nuclear shapes in heavy-ion reaction.
Quantum effects such as the Pauli principle and antisymmetrization of wave functions are also automatically taken into account in TDHF, which are essential for the manifestation of shell structures during the collision dynamics.
TDHF approach has many successful applications in the description of nuclear large amplitude collective motions,
as seen in recent applications to fusion~\cite{Umar2008_PRC77-064605,Guo2012_EPJWoC38-09003,
Umar2014_PRC89-034611,Washiyama2015_PRC91-064607,Godbey2017_PRC95-011601,Simenel2017_PRC95-031601,Guo2018_PLB782-401,
Guo2018_PRC98-064607}, quasifission~\cite{Golabek2009_PRL103-042701,Wakhle2014_PRL113-182502,Oberacker2014_PRC90-054605,Umar2015_PRC92-024621,
Umar2016_PRC94-024605,Sekizawa2016_PRC93-054616,Yu2017_SciChinaPMA60-092011,Guo2018_PRC98-064609,Zheng2018_PRC98-024622},
transfer reactions~\cite{Washiyama2009_PRC80-031602,Simenel2010_PRL105-192701,Scamps2013_PRC87-014605,Sekizawa2013_PRC88-014614,
Sekizawa2014_PRC90-064614,Sonika2015_PRC92-024603,Wang2016_PLB760-236,Sekizawa2016_PRC93-054616,Sekizawa2017_PRC96-041601,
Sekizawa2017_PRC96-014615,Roy2018_PRC97-034603,Jiang2018_CPC42-104105},
fission~\cite{Simenel2014_PRC89-031601,Scamps2015_PRC92-011602,Goddard2015_PRC92-054610,Goddard2016_PRC93-014620,Tanimura2017_PRL118-152501},
and deep inelastic collisions~\cite{Maruhn2006_PRC74-027601,Guo2007_PRC76-014601,Guo2008_PRC77-041301,Iwata2011_PRC84-014616,Dai2014_PRC90-044609,Dai2014_SciChinaPMA57-1618,
Stevenson2016_PRC93-054617,Guo2017_EPJWoC163-00021,Shi2017_NPR34-41,Umar2017_PRC96-024625}.
However, since the dynamical fluctuation and two-body dissipation are not included in TDHF, the fluctuations of collective variables are found to be considerably
underestimated~\cite{Davies1978_PRL41-632,Simenel2011_PRL106-112502}. To remedy the limitations in TDHF, the extension of theoretical framework beyond TDHF has been developed to include these effects, e.g., stochastic extension of TDHF theory~\cite{Ayik2009_PRC79-054606,Ayik2016_PRC94-044624,Tanimura2017_PRL118-152501,Ayik2017_PRC96-024611,Ayik2018_PRC97-054618,Yilmaz2018_PRC98-034604}, Balian-V$\acute{\rm e}$n$\acute{\rm e}$roni variational approach~\cite{Balian1984_PLB136-301,Simenel2011_PRL106-112502,Williams2018_PRL120-022501}, time-dependent density matrix approach ~\cite{Tohyama1987_PRC36-187,Tohyama2016_PRC93-034607}, time-dependent generator coordinate method~\cite{Berger1984_NPA428-23,Simenel2014_JPG41-094007}.

The main purpose of this work is to provide a predictive approach for production cross sections, and testify how and to what extent
the microscopic approach works well in multinucleon transfer reaction.
Multinucleon transfer dynamics may be affected by many variables, e.g., collision energy, deformation orientation, and shell structure of
the colliding nuclei. We will investigate the production cross sections and microscopic mechanism in multinucleon transfer reaction
by using the approach TDHF+GEMINI, which combines the microscopic TDHF with the state-of-art statistical model \texttt{GEMINI++}~\cite{GEMINI++_code,Charity1988_NPA483-371,Charity2010_PRC82-014610,Mancusi2010_PRC82-044610}, to take into
account both the multinucleon transfer dynamics and the secondary deexcitation process.
Recently, the combined method TDHF+GEMINI~\cite{Sekizawa2017_PRC96-041601,Sekizawa2017_PRC96-014615,Jiang2018_CPC42-104105} has been applied
to multinucleon transfer reactions and has described the production cross sections in an accuracy
comparable to the existing macroscopic models. In addition, we combine the TDHF theory with the statistical model HIVAP to study the reaction mechanism
of quasifission and fusion-fission dynamics in the reaction $^{48}$Ca+$^{239,244}$Pu, and find that the TDHF+HIVAP method well accounts for the experimental
observations on the isotopic dependence of fusion-evaporation cross sections~\cite{Guo2018_PRC98-064609}.
These results are rather remarkable given the fact that there is no adjustable parameters for the reaction dynamics in the microscopic TDHF method.

This article is organized as follows. In Sec.~\ref{theory} we recall the theoretical approaches of TDHF, particle-number projection, and \texttt{GEMINI++}
used for the calculations
of primary and secondary cross sections. Section~\ref{results} presents the production cross sections and microscopic mechanism in the multinuleon transfer reaction $^{58}$Ni+$^{124}$Sn by using TDHF+GEMINI approach.
We compare our results with experimental measurements and those from GRAZING model. A brief summary is given in Sec.~\ref{summary}.

\section{Theoretical Method TDHF+GEMINI}
\label{theory}

\subsection{Primary cross section}

In TDHF approach the many-body wave function is approximated as a single Slater determinant
\begin{equation}
\Psi({\bm r}_1\sigma_1 q_1,\cdots,{\bm r}_N\sigma_N q_N,t)=\frac{1}{\sqrt{N!}}\text{det}\{\phi_{\mathrm{\lambda}}({\bm r}_i\sigma_i q_i,t)\},
\end{equation}
where $\phi_{\rm{\lambda}}({\bm r}_i\sigma_i q_i,t)$ is the time-dependent single-particle states with spatial coordinate ${\bm r}_i$, spin $\sigma_i$, and isospin $q_i$ of $i$th ($i=1,\cdots,N$) nucleon.
This form is kept at all times in the dynamical evolution.
By taking the variation of time-dependent action with respect to the single-particle states, one may obtain a set of nonlinear
coupled TDHF equations in the multidimensional spacetime phase space
\begin{equation}
i\hbar\frac{\partial}{\partial t}\phi_{\mathrm{\lambda}}(\bm{r}\sigma q,t)=\hat h\phi_{\mathrm{\lambda}}(\bm{r}\sigma q,t),
\end{equation}
where $\hat h$ is the Hartree-Fock (HF) single-particle Hamiltonian. TDHF equation describes the time
evolution of single-particle wave functions in a mean field.

In TDHF description of heavy-ion collisions, the nucleon transfer happens when the projectile wave functions extend to the target spatial region and vice versa. As a result, TDHF Slater determinant after collision is not an eigenstate of particle-number operator, but a superposition of states with different particle numbers. By dividing the spatial region of total system into the subspace either the projectile-like fragments (PLF) or target-like fragments (TLF) located,
the particle-number operator in the subspace $V$ is defined as
\begin{equation}
\hat{N}_V = \int_V d\bm{r}\sum_{i=1}^{N_V}\delta(\bm{r}-{\bm r}_i)=\sum_{i=1}^{N_V}\Theta_V({\bm r}_i),
\end{equation}
where $\Theta_V({\bm r}_i)$=1 if ${\bm r}_i\in V$ and 0 elsewhere.
The particle-number projection operator~\cite{Simenel2010_PRL105-192701} in the subspace $V$
\begin{equation}
\hat{P}_n^V=\frac{1}{2\pi}\int_0^{2\pi}d\theta e^{i(n-\hat N_V)\theta}
\label{Eq:PNP_ope}
\end{equation}
projects the TDHF wave function onto the eigenstate with good particle number $n$.

After the particle-number projection, the state $\hat{P}_n^V|\Psi\rangle$ is the eigenstate of number operator $\hat{N}_V$
\begin{equation}
 \hat{N}_V\hat{P}_n^V|\Psi\rangle=n\hat{P}_n^V|\Psi\rangle.
\end{equation}
The expectation value of particle-number projection operator
\begin{equation}
 P_n=\langle \Psi|\hat{P}_n^V|\Psi\rangle,
\label{Eq:PNP_pro}
\end{equation}
is the distribution probability with $n$ nucleon number.
The probability for fragment with $Z$ protons and $N$ neutrons at a given impact parameter $b$ and incident energy $E_{\mathrm{c.m.}}$
\begin{equation}
P_{Z,N}(b, E_{\mathrm{c.m.}})=P_Z(b,E_{\mathrm{c.m.}})P_N(b,E_{\mathrm{c.m.}}),
\end{equation}
is a product of probabilities for protons $P_Z$ and neutrons $P_N$.
The cross section for a primary reaction product is then calculated by integrating the probability $P_{Z,N}$ over the impact parameters $b$
\begin{equation}
\sigma_{Z,N}(E_{\mathrm{c.m.}})=2\pi\int_{b_{\text{min}}}^{b_{\text{cut}}}b P_{Z,N}(b,E_{\mathrm{c.m.}})db,
\label{Eq:cs_pri}
\end{equation}
where $b_{\text{min}}$ is the minimum impact parameter for the binary reaction.
The cutoff impact parameter $b_{\text{cut}}$ should be chosen large enough in the numerical simulation to ensure outside which the transferred nucleon is negligibly small. It should be noted that this cross section is dependent on the deformation orientations of projectile and target nuclei. The total
cross section can be obtained by a proper integration over all the deformation orientations.

\subsection{Secondary cross section}

The primary reaction products are excited in the statistical nonequilibrium states and will undergo the deexcitation process including both
the evaporation of light particles and fission of heavy fragments. We employ the state-of-art statistical model \texttt{GEMINI++}~\cite{GEMINI++_code,Charity1988_NPA483-371,Charity2010_PRC82-014610,Mancusi2010_PRC82-044610}
to take into account this process and describe the production cross section of secondary reaction products.
The \texttt{GEMINI++} is the updated version of \texttt{GEMINI} based on the Monte-Carlo simulation.
This model describes the width of fission mass and charge distributions for heavy systems quite well~\cite{Charity1988_NPA483-371,Charity2010_PRC82-014610}.
The input parameters for \texttt{GEMINI++} code are the proton number $Z$, neutron number $N$, excitation energy $E^*_{Z, N}$, and angular momentum $J_{Z, N}$ of
the primary reaction product. The last two quantities for each primary product can be calculated, in principle, by the extended particle-number projection technique as done in Ref.~\cite{Sekizawa2014_PRC90-064614} for the light system $^{16}\text{O}+^{24}\text{O}$. But the large computational effort makes it difficult to estimate the two quantities for all primary products.
A simple but effective way is to evaluate the average values of these quantities as shown in Ref.~\cite{Sekizawa2017_PRC96-014615}.
The average excitation energy and angular momentum of the primary products can be directly  obtained from TDHF calculations.
The total excitation energy of the reaction system
\begin{equation}
E_{\mathrm{tot}}^*=E_{\mathrm{c.m.}}-\text{TKE}+Q,
\end{equation}
is expressed in terms of incident energy in center-of-mass (c.m.) frame $E_{\mathrm{c.m.}}$, total kinetic energy (TKE),
and transfer-channel dependent $Q$ value.
In TDHF, the TKE is calculated as the sum of kinetic energy of the fragments after the
separation and Coulomb potential energy assuming that the fragments are
pointlike charges. The transfer-channel dependent Q value is obtained by the latest experimental atomic mass tables AME2016~\cite{Huang2017_CPC41-030002,Wang2017_CPC41-030003} and the theoretical mass calculated by the finite-range droplet model FRDM(2012)~\cite{Moller2016_ADNDT109-1}. The total excitation energy is assumed to distribute over all the
transfer channels proportional to the mass of primary fragment
\begin{equation}
E_{Z,N}^*=\frac{Z+N}{A_1 + A_2 }E_{\text{tot}}^*,
\end{equation}
where $A_1$ and $A_2$ are the masses of projectile and target nuclei.
The average angular momentum is given by the expectation value of angular momentum operator
\begin{equation}
J_{Z,N}=\langle \Psi|\hat{J}_V|\Psi \rangle,
\end{equation}
with
\begin{equation}
\hat{J}_V=\sum_{i=1}^A\Theta_V({\bm r}_i)\left [({\bm r}_i-\bm{R}_{\text{c.m.}})\times \hat{\bm{p}}_i+\hat{\bm{s}}_i\right ],
\end{equation}
where $\bm{R}_{\text{c.m.}}$ is the c.m. position of the fragment, $\hat{\bm{p}}_i$ and $\hat{\bm{s}}_i$ are the single-particle momentum and spin operators, respectively.

For a given primary fragment specified by ($Z$, $N$, $E^*_{Z,N}$, $J_{Z,N}$), \texttt{GEMINI++} simulation follows the sequential binary-decays of all possible modes
for the compound nucleus.
Since \texttt{GEMINI++} is a statistical model based on Monte-Carlo algorithm, there may be  different binary-decay chains for a same set of input parameters.
In order to get the decay probability, we simulate the decay process $M_{\text{event}}$ times for a same set of input parameters.
We count the number of decay products composed of $Z^\prime$ protons and $N^\prime$ neutrons as $M_{Z^\prime,N^\prime}$ times.
The decay probability from the primary product with $(Z,N)$ to final product with $(Z^\prime, N^\prime)$ is given by
\begin{equation}
P_{\text{decay}}(E^\ast_{Z,N},J_{Z,N},Z,N;Z^\prime,N^\prime)=\frac{M_{Z^\prime,N^\prime}}{M_{\textrm{event}}}.
\label{Eq:decay_pro}
\end{equation}
It should be noted that the decay process with $P_{\text{decay}} \leq 1/{M_\text{event}}$ may not be taken into account due to
the limit of decay probability.
The probability for final products after the secondary deexcitation process
\begin{equation}
P^{(f)}_{Z^\prime,N^\prime}(b,E_{\textrm{c.m.}})=\sum_{Z\geq Z^\prime}\sum_{N\geq N^\prime} P_{Z,N}(b,E_{\textrm{c.m.}})P_{\text{decay}}(E^\ast_{Z,N},J_{Z,N},Z,N;Z^\prime,N^\prime),
\end{equation}
is a product of distribution probability and decay probability of primary fragments.
The production cross section for final product is evaluated as
\begin{equation}
\sigma^{(f)}_{Z^\prime,N^\prime}(E_{\textrm{c.m.}})=2\pi\int_{b_{\text{min}}}^{b_{\text{cut}}} b P^{(f)}_{Z^\prime,N^\prime}(b,E_{\textrm{c.m.}})db,
\end{equation}
which can be compared with the experimental measurement directly.

\section{Results and discussions}
\label{results}

We have tested the correctness of our code by comparing our results with those obtained from other codes.
We have reproduced accurately the transfer probability and the fluctuation of nucleon number for the reaction $^{16}\text{O}+^{208}\text{Pb}$
reported in Ref.~\cite{Simenel2010_PRL105-192701}. We also calculated the transfer cross sections in the reaction $^{40}\text{Ca}+^{124}\text{Sn}$
and found the results in Ref.~\cite{Sekizawa2013_PRC88-014614} are reproduced in an accurate precision by our code.

In the present work, we investigate the production cross sections and transfer mechanisms in the reaction $^{58}$Ni+$^{124}$Sn to testify
the applicability of microscopic approach TDHF+GEMINI in multinucleon transfer.
The production cross sections in multinucleon transfer reaction
$^{58}\text{Ni}+^{124}\text{Sn}$ have been measured in ANL experiment at
center-of-mass (c.m.) energies of 150, 153, 157,
and 160.6 MeV~\cite{Jiang1998_PRC57-2393}, at which the theoretical studies are performed. It should be noted that the experiment utilized the inverse kinematics method with a $^{124}\text{Sn}$ beam bombarding a $^{58}\text{Ni}$ target.
We employ the microscopic TDHF with Skyrme SLy5 parameter set~\cite{Chabanat1998_NPA635-231_NPA643-441} in the calculations, in which all the time-even and
time-odd terms in the mean-field Hamiltonian are included in our code. For details in the energy functional, see Refs.~\cite{Guo2018_PLB782-401,Guo2018_PRC98-064607,Guo2018_PRC98-064609}.
This force has been widely used in the fully three-dimensional TDHF calculations in heavy-ion collisions~\cite{
Umar2008_PRC77-064605,Sekizawa2013_PRC88-014614,Sekizawa2014_PRC90-064614,Wang2016_PLB760-236,Sekizawa2016_PRC93-054616,Sekizawa2017_PRC96-041601,Yu2017_SciChinaPMA60-092011, Sekizawa2017_PRC96-014615,Guo2018_PLB782-401,Guo2018_PRC98-064607, Guo2018_PRC98-064609,Stevenson2019_PPNP104-142,Sekizawa2019_FP7-20}.

In the numerical simulation, we first calculate the ground states of projectile and target nuclei by solving
the static HF equation on three-dimensional grid $24\times 24 \times 24$ ${\textrm{fm}}^3$. To avoid discrediting the local minimum as the ground state, we perform
the static HF calculations
with various initial potentials of spherical, prolate, oblate, and triaxial deformations.
This is necessary to find the true ground state, especially when the local minima are close to the HF ground state.
We find that both $^{58}$Ni and $^{124}$Sn show prolately deformed ground states with quadrupole deformation $\beta \sim $ 0.11 and 0.054, respectively, which are
same as the results from \texttt{Ev8} code~\cite{Ryssens2015_CPC187-175}. The ground state deformations in our calculations are also consistent
with the HF results of $^{58}$Ni~\cite{Sekizawa2013_PRC88-014614} and $^{124}$Sn~\cite{Oberacker2012_PRC85-034609}.
However, an oblate ground state with $\beta \sim $ 0.11 for $^{124}$Sn
has been obtained~\cite{Sekizawa2013_PRC88-014614,Washiyama2015_PRC91-064607} in HF calculations, implying a coexistence of prolate and oblate states with small energy difference.
This discrepancy in the ground state deformation of $^{124}$Sn may arise from the different treatment of numerical methods, e.g., the way to calculate the
derivatives of wave functions.
In a second step, we apply a boost
operator on the static wave functions to simulate TDHF time evolution with a time step $0.2~\mathrm{fm}/c$.
A numerical box $56\times 24\times 48$ ${\textrm{fm}}^3$ with a grid spacing $1$ fm has been used for the collision process.
The reaction is in $x$-$z$ plane and the collision axis is along $x$-axis.
The nucleus is assumed to move on a pure Coulomb trajectory until the initial separation distance 25~fm.
The choice of these parameters assures a good numerical accuracy for all the cases studied here.
Third, after TDHF simulation, we use the particle-number projection method to extract
the transfer probability and primary cross section from TDHF wave functions.
In the numerical calculation of transfer probability by using Eq.~(\ref{Eq:PNP_ope}), we divide the integration over the gauge angle into 300 uniform mesh,
and find that $P_n$ keeps almost unchanged as the increase of the number of mesh points.
In the calculation of primary cross section according to Eq.~(\ref{Eq:cs_pri}), we find that
the number of transferred nucleons is negligibly small at the cutoff impact parameter $b_{\textrm {cut}}>$ 9 fm.
Last is to evaluate the deexcitation process of primary products by using the state-of-art statistical model \texttt{GEMINI++}~\cite{GEMINI++_code,Charity1988_NPA483-371,Charity2010_PRC82-014610,Mancusi2010_PRC82-044610}. The cross
sections for secondary products can be used for a direct comparison with the experimental measurement. In present calculation, we employ the default parameter
setting in \texttt{GEMINI++} code, as did in most calculations~\cite{Sekizawa2017_PRC96-014615,Sekizawa2017_PRC96-041601,Jiang2018_CPC42-104105}.
We have also tested the dependence of decay probability on the Monte-Carlo simulation times $M_{\rm{event}}$, and found
that the decay probabilities are nearly identical for $M_{\rm{event}}=$1000 and 10000. Consequently,
we simulate the decay process 1000 times to evaluate the decay probability for present reaction system.

\begin{figure}
\includegraphics[width=0.46\textwidth]{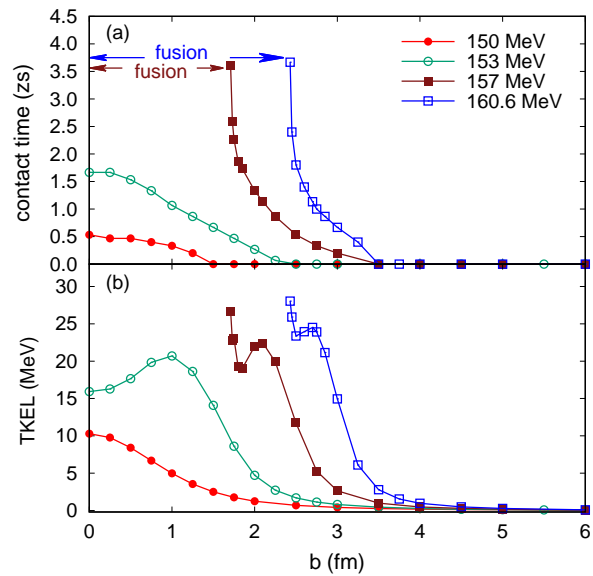}
\caption{(Color online) (a) contact time and (b) total kinetic energy loss (TKEL) as a function of impact parameter
for the tip collision $^{58}$Ni+$^{124}$Sn.
The four energies 150, 153, 157, and 160.6 MeV used in the calculations are those in ANL experiment~\cite{Jiang1998_PRC57-2393}.}
\label{Fig:time_TKEL}
\end{figure}

Since both $^{58}$Ni and $^{124}$Sn are prolately deformed in their ground states, the proper average over all the deformation orientations should be done
to provide the complete scenario of reaction dynamics. However, due to the large computational cost in the microscopic TDHF calculations, we perform the
calculations at two extreme orientations (tip and side) in present work. The so-called tip (side) orientation refers that the deformation axis of nucleus
is initially set parallel (perpendicular) to the collision axis. Here and thereafter, the tip (side) collision is phrased when both $^{58}$Ni and $^{124}$Sn
are initially set to be the tip (side) orientation. We calculate the TDHF capture barrier for two extreme collisions (tip and side).
For the tip collision, the barrier is found to be 153.8 MeV, while the side collision results in a significantly
higher barrier of 160.6 MeV, as expected. The energies used in ANL experiments are quite close to Coulomb barrier.

It should be noted that both $^{58}$Ni and $^{124}$Sn present the spherical ground states with the inclusion of pairing correlations in HF+BCS calculations.
The possible transfer of a correlated pair or a cluster of nucleons may play a role in multinucleon transfer reaction. In recent years the inclusion of pairing
correlations becomes possible in the microscopic simulation of collision dynamics~\cite{Scamps2013_PRC87-014605,Ebata2015_JPSCP6-020056,Hashimoto2016_PRC94-014610,Magierski2017_PRL119-042501,Sekizawa2017_EPJWoC163-00051}.
However, the influence of pairing correlations in
multinucleon transfer reaction is still an open question. In the present work, we focus on the TDHF studies of multinucleon transfer, and the inclusion of pairing
correlations will be the subject of future works.

\begin{figure}
\includegraphics[width=0.46\textwidth]{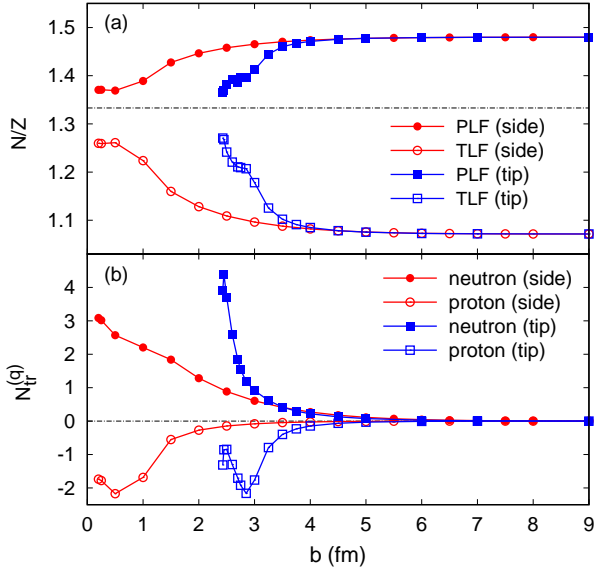}
\caption{(Color online) (a) ratio of neutron to proton $N/Z$ for PLF (solid symbol) and TLF (open symbol), and the horizontal line
for the equilibrium value of total system, and (b) transferred nucleon number for neutrons (solid symbol) and protons (open symbol).
The side and tip collisions $^{58}$Ni+$^{124}$Sn at the energy 160.6 MeV are denoted by circle and square symbols, respectively.}
\label{Fig:particle}
\end{figure}

We first consider the energy dependence of transfer dynamics in the tip collision of $^{58}$Ni+$^{124}$Sn.
The contact time (a) and total kinetic energy loss (TKEL) (b) as a function of impact parameter $b$ are shown
in Fig.~\ref{Fig:time_TKEL}. The four energies 150 MeV (solid circle), 153 MeV (open circle), 157 MeV (solid square), and 160.6 MeV (open  square)
are used in the calculations, at which the experimental measurement has been performed in ANL~\cite{Jiang1998_PRC57-2393}.
Here the contact time is calculated as the time interval in which the lowest density on the line between the mass centers of the two fragments exceeds half of the nuclear
saturation density $\rho_{0}/2=0.08~\mathrm{{fm}^{-3}}$, as used in Refs.~\cite{Oberacker2014_PRC90-054605,Umar2016_PRC94-024605,Yu2017_SciChinaPMA60-092011,Guo2018_PRC98-064609}.
We observe that the contact time presents a rapid decrease as the increase of impact parameter, which is a typical change
from the central to peripheral collisions. In the more central collision, the elongation
of the dinuclear system is much slower and the compact configuration with mononuclear shape remains much longer, which are expected to lead to a longer contact time.
At higher energies 157 and 160.6 MeV, fusion happens at the impact parameter smaller than 1.71 fm and 2.43 fm, in which the collective kinetic energy is entirely
converted into the internal excitation of a well-defined compound nucleus. Since the product of proton numbers of target and projectile nuclei
$Z_{\textrm P}Z_{\textrm T}=1400$ is smaller than the critical value 1600, the quasifission dynamics is not expected in this reaction.
Hence the process is considered as fusion reaction leading to the formation of compound nucleus, when the contact time of two colliding nuclei is larger than 4000 fm/c.

\begin{figure}
\includegraphics[width=0.47\textwidth]{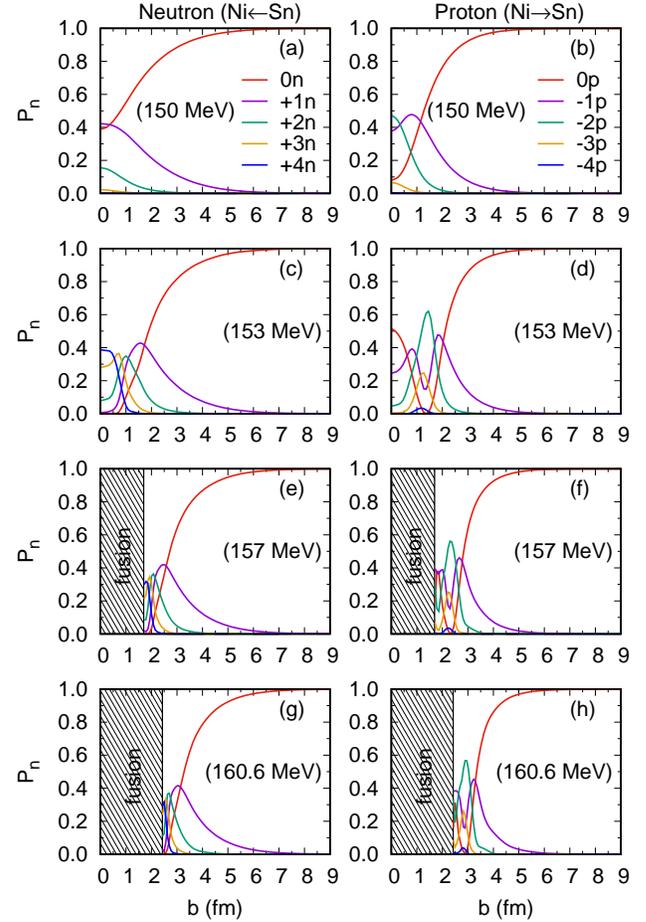}
\caption{(Color online) Transfer probabilities of neutron pickup (left panels) and proton removal (right panels) channels with respect to the target nucleus $^{58}$Ni.
The results are for the tip collision at the energies 150 MeV (top panels), 153 and 157 MeV (middle panels), and 160.6 MeV (bottom panels).
The shaded region corresponds to the fusion reaction.}
\label{Fig:trans_proba_tip}
\end{figure}

In Fig.~\ref{Fig:time_TKEL}(b), TKEL is calculated as the energy difference of incident energy and TKE, $\textrm{TKEL}=E_{\textrm{c.m.}}-\textrm{TKE}$.
TKEL increases with the incident energy due to the fact that more nucleons are excited at higher energies, leading to more energies dissipated from collective energy to intrinsic excitation and larger loss of TKE.
TKEL decreases as a function of impact parameter until
nearly zero at $b=$5 fm, which corresponds to the zero contact time and a quasielastic reaction.
We also observe a plateau region of TKEL at around $b=$1.0 fm for 153 MeV, 2.1 fm for 157 MeV, and 2.7 fm for 160.6 MeV.
The time evolution of density distribution
reveals that the plateau pattern in TKEL arises from the dinuclear property of
the neck whose formation is observed when the TKEL becomes substantial.

We next look at the dependence of transfer dynamics on the deformation orientation of colliding partners. In Fig.~\ref{Fig:particle},
the ratio of neutron to proton $N/Z$ for PLF (solid symbol) and TLF (open symbol) (a), and
transferred nucleon number for neutrons (solid symbol) and protons (open symbol) (b)
are shown as a function of impact parameter. The results are for the side (circle symbol) and tip collisions (square symbol) at the energy 160.6 MeV.
We find that both the ratio $N/Z$ and the transferred nucleon number in the side collision show more flat distribution as a function of impact parameter as compared to the tip collision, indicating an overall dependence of deformation orientations of projectile and target nuclei.
The ratio $N/Z$ is 1.07 for the target nucleus $^{58}$Ni, 1.48 for the projectile $^{124}$Sn, and 1.33 for the equilibrium value of the total system.
Due to the large $N/Z$ asymmetry between projectile and target nuclei, the nucleons are expected to transfer toward the direction of charge
equilibration to reduce the $N/Z$ asymmetry. Namely, the protons transfer from $^{58}$Ni to $^{124}$Sn, while the transfer for neutrons is
in opposite direction from $^{124}$Sn to $^{58}$Ni.
As shown in the upper panel of Fig.~\ref{Fig:particle}, TDHF simulation indeed gives a scenario of such a transfer mechanism.
The ratio $N/Z$ is nearly same as
the initial value of individual nucleus at $b>5$ fm , and then approaches to the equilibration value of total system (the horizontal line) as the decrease of impact parameter.
At the impact parameter $b<0.21$ fm for the side collision and $b<2.43$ fm for the tip collision, the fusion reaction happens.

\begin{figure}
\includegraphics[width=0.47\textwidth]{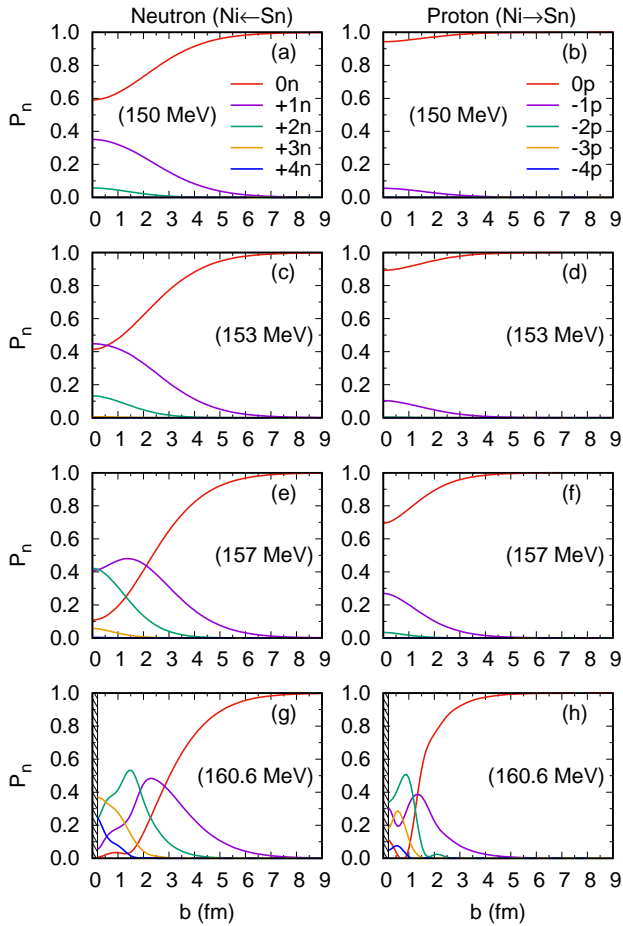}\\
\caption{(Color online) Same as Fig.~\ref{Fig:trans_proba_tip}, except for the side collision.}
\label{Fig:trans_proba_side}
\end{figure}

At energies close to the Coulomb barrier, the nucleon emitted into continuum in the breakup process is negligibly small as shown in Ref.~\cite{Sekizawa2013_PRC88-014614}. As a result, the transferred nucleon number from target to projectile is evaluated as $N^{(q)}_{\textrm{tr}}=N^{(q)}_{\textrm {TLF}}-N^{(q)}_{\textrm {T}}$  with isospin
index $q$. As shown in the lower panel of Fig.~\ref{Fig:particle}, the positive and negative values of transferred nucleon number correspond to the neutron pickup and proton removal channels with
respect to the target nucleus $^{58}$Ni, respectively. At the relatively small impact parameter, a large number of nucleons are transferred.
As the increase of impact parameter, the number of transferred nucleons decreases until nearly zero at $b>6$ fm.

The transfer probability for each reaction channel is extracted from the final TDHF wave functions by using the particle-number projection method
according to Eq.~(\ref{Eq:PNP_pro}). We find that the transfer probabilities toward the direction of charge equilibrium are at least one order of magnitude larger than those in opposite transfer due to the large $N/Z$ asymmetry between projectile and target nuclei. Hence, we show only the transfer probabilities toward the direction of charge equilibrium in Fig.~\ref{Fig:trans_proba_tip} for the tip collision $^{58}$Ni+$^{124}$Sn.
The left (right) panels correspond to the transfer that neutrons (protons) are added to (removed from) the target nucleus $^{58}$Ni.
The shaded region denotes the fusion reaction.
We perform the calculations at the energies 150 MeV (top panels), 153 and 157 MeV (middle panels), and 160.6
MeV (bottom panels).
As the increase of incident energies, we see that more nucleons are transferred with non-zero probabilities, because at higher energy the contact time
is long enough for the transfer of more nucleons.
For the $(0n)$ and $(0p)$ channels the probabilities increase with the impact parameter until nearly equal to one at $b>7$ fm, while the transfer in $(xn)$ and $(-xp)$
$(x\geq 1)$ channels shows a peak at the relatively small impact parameter and then decreases to nearly zero at $b>7$ fm, indicating a transition from multinucleon transfer to quasielastic reactions.
For the more central collisions, the behavior of multinucleon transfer shows complicated dependence on the impact parameter and incident energy.

\begin{figure}
\includegraphics[width=0.48\textwidth]{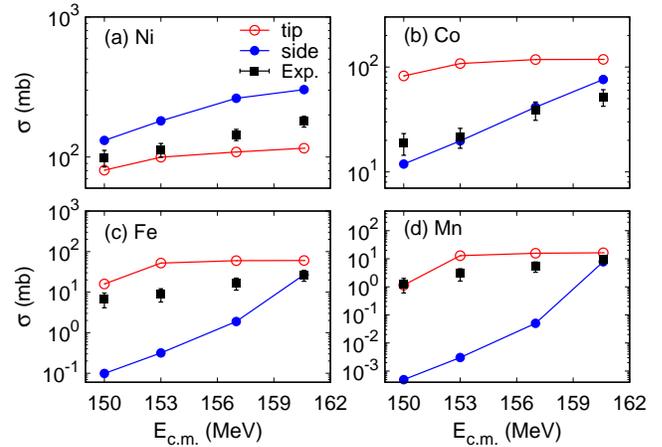}\\
\caption{(Color online) Energy dependence of the total cross sections integrated over all the neutron pickup channels in the reaction $^{58}$Ni+$^{124}$Sn.
The results for the tip and side collisions are denoted by the open and solid circles, respectively. The experimental data with errorbars~\cite{Jiang1998_PRC57-2393} is also included for comparison.}
\label{Fig:cross_section_int}
\end{figure}

The transfer probabilities for the side collision are shown in Fig.~\ref{Fig:trans_proba_side}. We find that less nucleons
are transferred as compared to the tip collision because of the higher Coulomb barrier in the side collision.
The transfer probabilities in the $(0n)$ and $(0p)$ channels at the relatively small impact parameter are systematically larger than those
in the tip collision, while the transfer in $(xn)$ and $(-xp)$ $(x\geq 1)$ channels has small probabilities in the side collision,
satisfying the condition that the sum of transfer probabilities over all reaction channels equals to one.
This distinct behavior of transfer probabilities may lead to the larger cross sections in the $(0p)$ channel
for the side collision as compared to the tip case, which will be shown in Fig.~\ref{Fig:cross_section_int}.

\begin{figure}
\includegraphics[width=0.48\textwidth]{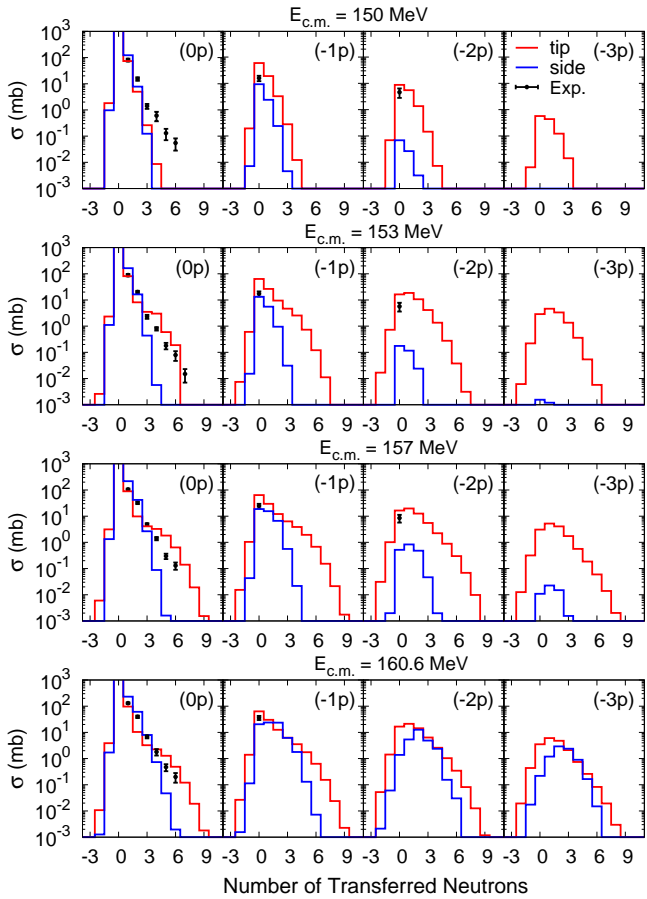}\\
\caption{(Color online) Production cross sections for secondary fragments obtained from TDHF+GEMINI together with the experimental data~\cite{Jiang1998_PRC57-2393} in the reaction $^{58}$Ni+$^{124}$Sn. The calculations are performed for the tip and side collisions
at the energies 150 MeV (top panels), 153 and 157 MeV (middle panels), and 160.6 MeV (bottom panels).}
\label{Fig:cross_section}
\end{figure}

Since the incident energies are close to the Coulomb barrier, the primary fragments with the low excitation energies will undergo the weak
deexcitation process. We find that both the emission of light particles and fission of heavy primary fragments are insignificant that
the cross sections for primary and secondary products are quite close to each other. Hence, we show only the production
cross sections of secondary fragments at the energies 150 MeV (top panels), 153 and 157 MeV (middle panels), and 160.6 MeV (bottom panels) in Fig.~\ref{Fig:cross_section}.
The results obtained from TDHF+GEMINI calculations for the tip and side collisions are denoted by the red and blue histograms, respectively.
For comparison, the experimental data with errorbars~\cite{Jiang1998_PRC57-2393} is also included by black solid circles.
In principle, the tip and side collisions correspond to the two extreme (upper and lower) limit of the cross sections.
An accurate value of theoretical cross section can be obtained by the proper integration over all the deformation orientations,
which should locate in between the tip and side collisions.

We observe that the cross sections in the tip collision extend in a wide distribution as a function of the number of transferred neutrons as compared to the
side collision. This is consistent with the observation that more nucleons are transferred in the tip collision, as shown in Figs.~\ref{Fig:trans_proba_tip}
and \ref{Fig:trans_proba_side}.
At the higher energies 153, 157, and 160.6 MeV, the theoretical results obtained from TDHF+GEMINI well reproduce the experimental data for all the reaction channels. However, at the sub-barrier energy 150 MeV, the theoretical cross sections
in the $(0p)$ panel gradually deviate from the experimental measurements as the increase of the number of transferred neutrons, although the theoretical calculation in the most dominant channel $(0p,1n)$ agrees well with the measurement.
The discrepancy between the theoretical and experimental results at sub-barrier energy arises from the limitations of single mean-field approximation in TDHF approach. For example, the missing of many-body correlations such as the dynamical fluctuation and internucleon pairing
correlations in TDHF is partly responsible for the discrepancy. In addition, the realistic potential in multinucleon transfer process should depend on
the transfer channel. However,
the single mean-field potential in TDHF is not transfer-channel dependent, which describes the average value of probability distribution. Hence, the peak
position in the distribution corresponds to the average number of transferred nucleons, leading to the observed deviation between theoretical results
and experimental data at sub-barrier energy.
To perfectly reproduce the experimental cross sections at sub-barrier energy
in multinucleon transfer reaction, the description beyond the standard mean-field theory should be performed.

\begin{figure}
\includegraphics[width=0.48\textwidth]{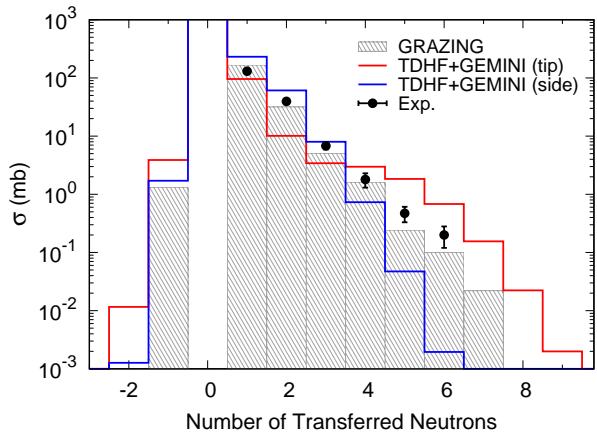}\\
\caption{(Color online) Production cross sections of secondary fragments in $(0p)$ channel at the energy 160.6 MeV for the
reaction $^{58}$Ni+$^{124}$Sn. The results obtained from TDHF+GEMINI for the tip and side collisions are shown by red and blue
histograms, respectively. The GRAZING results and experimental data~\cite{Jiang1998_PRC57-2393} are also included for comparison.}
\label{Fig:cross_section_GRAZING}
\end{figure}

For an overall description of the transfer dynamics, the energy dependence of the total cross sections integrated over all the neutron pickup channels is compared with the experimental measurement~\cite{Jiang1998_PRC57-2393} in Fig.~\ref{Fig:cross_section_int}.
The results for the tip and side collisions are denoted by the open and solid circles, respectively. We find that the experimental data basically locates in between the
two extreme theoretical values for the tip and side collisions. The agreement between the theoretical
results and experimental data is observed to be quite good for all the energies.
We also find that the cross sections for the proton stripping reactions to produce Co, Fe and Mn isotopes are larger in the tip collision than
the side collision. However, for the pure neutron transfer reactions to produce Ni isotopes, the cross sections present an opposite behavior that
the cross sections are large in the side collision. This opposite trend is attributed to
the large probabilities in the $(0p)$ channel for the side collision, which are shown in the relevant discussion of Fig.~\ref{Fig:trans_proba_side}.

For comparison with the existing model, the production cross sections in $(0p)$ channel at the energy 160.6 MeV from
TDHF+GEMINI calculations are compared with GRAZING results and experimental data in Fig.~\ref{Fig:cross_section_GRAZING}.
The TDHF+GEMINI results for the tip and side collisions are shown by red and blue
histograms, respectively.
The GRAZING results and experimental data are taken from Ref.~\cite{Jiang1998_PRC57-2393}.
We find that both the TDHF+GEMINI and GRAZING results are in good agreement with the experimental data. The GRAZING results locate in between the two extreme values of TDHF+GEMINI calculations. It should be noted that the GRAZING calculation does not take into account the deformation effects. The microscopic TDHF+GEMINI calculations give the quantitatively good description as the semiclassical GRAZING model. This is rather impressive because there exists no adjustable parameters for the reaction dynamics
in the microscopic TDHF method.

\section{SUMMARY}
\label{summary}

In present paper, we combine the microscopic TDHF approach with the state-of-art statistical model \texttt{GEMINI++} to investigate the multinucleon transfer dynamics
in the reaction $^{58}$Ni+$^{124}$Sn.
The TDHF+GEMINI approach takes into account both the multinucleon transfer dynamics and the secondary deexcitation process. After TDHF dynamical simulations,
the particle-number projection method is used to extract the transfer probability for each reaction channel. We investigate the dependence of transfer dynamics
on the incident energy and deformation orientations of projectile and target nuclei.
The transfer probability toward the direction of charge equilibration is observed
to be at least one order of magnitude larger than those in opposite transfer due to the large asymmetry $N/Z$ between projectile and target nuclei.
More nucleons are observed to transfer in the tip collision as compared to the side collision.
The production cross sections at above-barrier energies obtained from TDHF+GEMINI well reproduce the experimental measurement performed in ANL for all the reaction
channels. However, due to the limitations of single mean-field approximation in the microscopic TDHF theory, the cross sections at sub-barrier energy
gradually underestimate the
experimental data as the increase of the number of transferred neutrons.
For an overall description of the transfer dynamics, the energy dependence of the total cross sections integrated over all the neutron pickup channels are compared with the experimental data and the agreement is quite good for all the energies.
We also compare our results with those from GRAZING model, and find that our results give quantitatively good description as the macroscopic GRAZING model. This is impressive since
there is no adjustable parameters for the reaction dynamics in the microscopic TDHF calculations. These studies demonstrate the feasibility and success of TDHF+GEMINI approach in the microscopic reaction mechanism of multinucleon transfer dyanmics.

\begin{acknowledgments}
We thank Kazuyuki Sekizawa for helpful discussions and valuable comments on this article .
This work is partly supported by NSF of China (Grants No. 11575189 and No. 11175252), NSFC-JSPS International
Cooperation Program (Grant No. 11711540016), and Presidential Fund of UCAS. The computations in present work have been performed on the High-Performance Computing Clusters of SKLTP/ITP-CAS and Tianhe-1A supercomputer located in the
Chinese National Supercomputer Center in Tianjin.
\end{acknowledgments}

\bibliographystyle{apsrev4-1}
\bibliography{ref}
\end{document}